\documentclass[12pt]{article}

\usepackage{psfig}

\def\etal{{et\,al.}}
\def\msun{M$_{\odot}$}
\def\rsun{R$_{\odot}$}
\def\mdot{$\dot M$}
\def\degs{\ifmmode ^{\circ}\else$^{\circ}$\fi}
\def\amin{\ifmmode ^{\prime}\else$^{\prime}$\fi}
\def\asec{\ifmmode ^{\prime\prime}\else$^{\prime\prime}$\fi}
\def\farcs{\hbox{$.\!\!^{\prime\prime}$}}  

\def\shead #1{\centerline{\bf #1}}
\def\sshead #1{\centerline{\it #1}}
\def\amin{\ifmmode ^{\prime}\else$^{\prime}$\fi}
\newbox\grsign \setbox\grsign=\hbox{$>$}
\newdimen\grdimen \grdimen=\ht\grsign
\newbox\laxbox \newbox\gaxbox
\setbox\gaxbox=\hbox{\raise.5ex\hbox{$>$}\llap
     {\lower.5ex\hbox{$\sim$}}}\ht1=\grdimen\dp1=0pt
\setbox\laxbox=\hbox{\raise.5ex\hbox{$<$}\llap
     {\lower.5ex\hbox{$\sim$}}}\ht2=\grdimen\dp2=0pt
\def\gax{$\mathrel{\copy\gaxbox}$}

\def\grs{GRS 1915+105}

\begin{document}

\shead{An unusually massive stellar black hole in the Galaxy}

\bigskip

\shead{J. Greiner}
\sshead{Astrophysical Institute Potsdam, 14482 Potsdam, Germany}
\shead{J.G. Cuby}
\sshead{ESO, Alonso de C\'ordova 3107, Santiago 19, Chile}
\shead{M.J. McCaughrean}
\sshead{Astrophysical Institute Potsdam, 14482 Potsdam, Germany}

\bigskip
\bigskip

\noindent {\bf 
GRS 1915+105 belongs to the small group of galactic X-ray binaries
dubbed microquasars \cite{mr98, gre00}, which show
sporadic ejection of matter at apparently superluminal speeds. 
Knowing the basic system parameters of GRS 1915+105 is not only a prerequisite
of understanding the ingredients of jet formation, but also provides
the link to many other astrophysical objects which exhibit jets,
in particular extragalactic objects.
Based on its large X-ray luminosity during high-states of 
7$\times$10$^{39}$ erg/s \cite{grm99} and the
interpretation of the X-ray timing properties \cite{mrg97}, 
a mass of $\sim$10--30 \msun\ has been suggested for the accreting compact 
object in \grs. It is also one of only two galactic binary sources which 
are thought to contain a maximally spinning black hole \cite{zcc97}.
Here we report a measurement of the orbital period and mass function
of GRS 1915+105 which allow us to deduce a mass of 14 $\pm$ 4 \msun\
for its black hole.
This large mass provides a challenge for the black hole formation scenarios 
in binaries, since 
black holes with masses above 5--7 \msun\ 
are hard to explain \cite{blb99, wel99, blt01, kal01}.
Also, the mass estimate allows us to understand the unique X-ray variability 
of GRS 1915+105 as being due to instabilities of a radiation-pressure
dominated disk radiating near the Eddington limit.
Finally, several models are constrained which
relate observable X-ray properties to the spin of 
black holes in microquasars. Once further calibrated, these relations
may soon turn into a valuable tool to study relativistic effects in strong
gravitational fields.
}

\bigskip

\grs\ is located in the galactic plane at a distance of $\sim$11--12 kpc 
\cite{mr94, fen99}
and suffers a large extinction of 25--30 mag
in the visual band. Spectroscopic observations
in the near-infrared H and K bands 
were successful in identifying absorption features
from the atmosphere of the companion (mass donating star) in the
GRS 1915+105 binary \cite{gcm01}, and the detection of $^{12}$CO and
$^{13}$CO band heads plus a few metallic absorption lines suggested
a K-M spectral type and luminosity class III (giant).

The presence of these band-head features led us to carry out follow-up
medium-resolution spectroscopy in the 2.39--2.41 $\mu$m wavelength range 
using the VLT-Antu equipped with ISAAC,
between April and August 2000 (Fig. 1).
Radial velocities were measured for the 16 individual spectra
by cross-correlation of the major CO band heads, 
and a period analysis carried out (Fig. 2). 
The periodogram shows
a clear peak at a period of 33.5 days (top panel) which we interpret
as the orbital period $P_{\rm orb}$ of the binary system. 
The velocity amplitude is measured to be $K_{\rm d}$ = 140$\pm$15 km/s
(lower panel).
Figure 1 shows that the infrared flux is dominated by light from the
accretion flow or jet, rather than from the secondary star.  There is thus a
possibility that phase-dependent changes in the continuum near the
absorption features may result in an additional source of systematic error
in the measured value of $K_{\rm d}$.
The measured parameters allow us to determine the mass function $f(M)$, 
i.e. the observational lower limit to the mass of the compact object

\begin{equation}
  f(M) \equiv {(M_{\rm c} \, {\rm sin}\, i)^3 \over (M_{\rm c} + M_{\rm d})^2} 
            = {P_{\rm orb} \, K_{\rm d}^3 \over 2 \, \pi \, G}
            = 9.5 \pm 3.0 \, M_{\odot} .
\end{equation}

In order to determine the true mass of the black hole, estimates of the
donor mass $M_{\rm d}$ and the orbital inclination $i$ are required.
The K-M III classification, at first approximation, implies a mass of  
$M_{\rm d}$ = 1.2$\pm$0.2 \msun\ for the donor \cite{gcm01}.
Because of the high mass-loss of the donor (which is needed
to explain the large X-ray luminosity), the donor is most certainly
less luminous than a non-interacting star of the same spectral type.
This in turn would imply a larger donor mass, thus making the black hole mass
estimate (see below) a lower limit when using $M_{\rm d}$ = 1.2 \msun.
The orbital inclination of the GRS 1915+105 binary can be deduced from the
orientation of the jet, which in turn is derived from the brightness and the 
velocities of both the approaching and the receding 
blobs \cite{fen99, mr94}. This angle of $\approx$ 70\degs\ $\pm$ 2\degs\
was observed to be constant over several years, indicating no measurable
precession, and justifies us assuming that the
jet is perpendicular to the accretion disk and orbital plane.
Thus, knowing the inclination $i$ and a lower limit of the donor mass, we can 
solve Eq. 1
for the mass of the accreting compact object (Fig. \ref{mass}), finding 
$M_{\rm c} = 14 \pm 4$ \msun. 
Table \ref{res} summarises all orbital parameters of GRS 1915+105. 
Even after accounting for the 
relatively large error dominated by the  
determination of the velocity amplitude $K_{\rm d}$, GRS 1915+105 is the 
galactic low-mass X-ray binary with the largest mass function and the largest 
mass of its compact object. Previous record holders were 
V404 Cyg ($f(M) = 6.07 \pm 0.05$, $M_{\rm c} = 7 - 10$ \msun\ \cite{sbc96})
and XTE J1118+480 ($f(M) = 6.00 \pm 0.36$, $M_{\rm c} = 6.5 - 10$ \msun\ 
\cite{mgc01}).

The knowledge of the mass of the black hole in GRS 1915+105
has several implications for our understanding of the physics
in microquasars,
as well as some broader astrophysical concepts.
Most importantly, the formation of a 14 \msun\ black hole in a low-mass 
binary poses an interesting challenge for binary evolution models.
Stellar evolution of stars in a binary system proceeds differently from 
single star evolution primarily due to the mass transfer between the system 
components and/or common-envelope phases.
There are, in general, two different paths for the black hole formation
in a binary system. First, the progenitor system could be wide, and 
during the common envelope phase the low-mass (main sequence) star of 
$\sim$1 \msun\ will spiral into the envelope of the massive giant
(progenitor of the black hole), causing the orbit to shrink 
\cite{kal01, kal99}.
Based on our measured system parameters (Table 1), the deduced orbital 
separation of the binary components in GRS 1915+105 is 108$\pm$4 \rsun.
Thus, 
orbital contraction through a common-envelope phase 
caused by the expansion of the massive progenitor to typically \gax 1000 \rsun,
is conceivable for GRS 1915+105.
Second, the evolution could start with a progenitor system smaller
than today, provided the binary component interaction is delayed until
after He burning has ceased \cite{blb99}. In this case, the time between the
wind phase and the core-collapse is short, and black hole masses in the
5--10 \msun\ range are plausible when the initial He-star
progenitor is in the 10--25 \msun\ range,
corresponding to initial primaries with 25--45 \msun\ \cite{blt01, kal01}.
How much mass is finally lost
depends on the evolution of the two progenitor star radii,
and it remains to be shown whether black hole masses above 10 \msun\ can
be achieved.
In order to produce even higher black hole masses, the progenitor 
might have been 
a massive Wolf-Rayet star. However, Wolf-Rayet stars have a much 
larger wind-loss rate, and it is therefore unclear whether higher progenitor
masses indeed will lead to higher final black hole masses. 
An alternative way of producing high-mass (\gax\ 10 \msun) black holes 
may be to invoke hierarchical triples as progenitors \cite{egv86}.

Turning  to the details of accretion disk physics and microquasar
phenomenology, the knowledge of the mass of the accretor in  GRS 1915+105
also yields insight into the rapid and large-amplitude X-ray variability 
seen uniquely in this source \cite{gmr96}, and which 
occurs near or even above the Eddington limit $\dot M_{\rm Edd}$.
Such high accretion rates are never reached by  
other canonical black hole transients (e.g. GRO J1655-40)
which usually operate in the 0.1-0.2 $\dot L_{\rm Edd}$ range, at which
their accretion disks are likely gas pressure dominated, and thus viscously 
and thermally
stable. The uniquely high \mdot/$\dot M_{\rm Edd}$ ratio in GRS 1915+105
suggests that its inner accretion disk is radiation pressure 
dominated, which in turn makes the inner disk quasi-spherical and
thermally unstable.
This property provides a potential clue for the spectacular and unique
X-ray variability in GRS 1915+105 \cite{gmr96}.
While it is tempting to conclude that jet ejection occurs because the
black hole can not accept this copious supply of matter, it is
important to remember that jet ejection occurs also in these other sources,
(e.g. at 0.2 $\dot M_{\rm Edd}$ in GRO J1655-40),
and thus near/super-Eddington accretion cannot be the determining factor
for relativistic jets.

Finally, if the black hole mass in GRS 1915+105 is indeed no larger than 
18 \msun\ (Fig. 3), we can
place constraints on the black hole spin in GRS 1915+105 and GRO J1655-40.
Previously, information on the black hole spin has been deduced from two
completely different sources First, accretion disks around a (prograde) 
spinning black hole extend farther down towards the black hole, and thus
allow the temperature of the disk to be higher. Since both 
GRS 1915+105 and GRO J1655-40 exhibit a thermal component in their X-ray
spectra which is unprecedently high compared to all other black hole
transients (during outbursts), it has been argued that this is due to their
black hole spin, while the majority of
black hole transients have non-rotating black holes \cite{zcc97}.
Second, several black hole binaries, including GRS 1915+105 and GRO J1655-40,
show nearly-stable quasi-periodic oscillations (QPOs) in their X-ray emission.
The frequencies $f$ for these QPOs are 300 Hz in GRO J1655-40 \cite{rmm99}
and 67 Hz in GRS 1915+105 \cite{mrg97}.
Most of the models proposed to explain these QPOs either rely or depend
on the spin of the accreting black hole.
The knowledge of the black hole mass of
GRS 1915+105 makes the deduction of the black hole spin of \cite{zcc97}
inconsistent with any of these four models on the origin of QPOs.
(1) If associated with the Keplerian motion at the
  last stable orbit around a (non-rotating) black hole according to
  the simple relation
  $f$ (kHz)$ = 2.2 / M_{\rm BH}$ (\msun), it gives 
  a surprising agreement with the optically determined mass for GRO J1655-40,
  but is off by a factor of 2 for GRS 1915+105, i.e. the QPO frequency
  does not scale linearly with the mass of the black hole.
(2) If associated with the trapped g-mode (diskoseismic) oscillations near
  the inner edge of the accretion disk \cite{okf87, nwb97}
  would require a nearly maximally spinning black hole in
  GRO J1655-40, and a non-spinning black hole in GRS 1915+105.
(3) Similarly, if associated with the relativistic dragging of inertial frames
  around a spinning black hole \cite{czc98}
  which would cause the
  accretion disk to precess, the implied specific angular momentum (spin) of 
  the black hole in
  GRS 1915+105 would be $a \sim 0.8$, thus considerably lower than
  the $a \sim 0.95$ deduced for GRO J1655-40. 
The implications of both of these models are in conflict
  with the nearly identical accretion disk temperatures for both sources
  which in turn requires a larger spin
  for GRS 1915+105 \cite{zcc97}.
(4) If associated with oscillations related to a centrifugal barrier 
 in the inner part of the accretion disk \cite{tlm98},
 the product of QPO frequency and black hole mass is predicted to be
 proportional to the accretion rate, implying that the accretion rate
 in GRO J1655-40 should be a factor $\sim$10 larger than in GRS 1915+105.
 This is certainly not the case. 

Thus, none of these four models provides a satisfactory
solution if one adopts the interpretation that the high accretion disk 
temperatures are a measure of the black hole spin \cite{zcc97}.
If this latter interpretation is dropped, however, and the spin 
becomes a free parameter, the first three models could be applicable.
It should be noted that the applicability of the applied disk model 
to deduce accurate accretion disk temperatures has been also questioned 
on other grounds \cite{scr99, mfr99},
and that there also exist alternative models, so-called slim disk 
models, which can reproduce high-temperature disks also for non-rotating 
black holes \cite{wft00}.

{}

\noindent{\it Acknowledgements:}
This work is
based on observations collected at the European Southern Obser\-vatory,
Chile under proposal ESO No. 65.H-0422.

\newpage

\begin{table}
\caption{Spectroscopic Orbital Parameters of GRS 1915+105}
\label{res}
\begin{center}
\begin{tabular}{ll}
      \hline
      \noalign{\smallskip}
      Parameter     & Result \\
      \noalign{\smallskip}
      \hline
      \noalign{\smallskip}
   $T_0$ (UT)$^{(1)}$           & 2000 May 02 00:00 \\
   $T_0$ (Heliocentric)$^{(1)}$ & HJD 2\,451\,666.5$\pm$1.5 \\
   $\gamma$ (km/s)              & --3$\pm$10 \\
   $K_{\rm d}$ (km/s)           & 140$\pm$15 \\
   $P_{\rm orb}$ (days)         & 33.5$\pm$1.5 \\
   $f(M)$ (\msun)               & 9.5$\pm$3.0 \\
   $M_{\rm d}$ (\msun)          & 1.2$\pm$0.2 \\
   $M_{\rm c}$ (\msun)$^{(2)}$  & 14$\pm$4 \\
 \noalign{\smallskip}
 \hline
 \end{tabular}
 \end{center}

\noindent{
   $^{(1)}$ Time of blue-to-red crossing. \\
   $^{(2)}$ Using an inclination angle of $i$ = 70$\pm$2 degrees 
            \cite{fen99, mr94} (see caption of Fig. 3).
  }
 \end{table}

\newpage

\begin{figure}
    \vbox{\psfig{figure=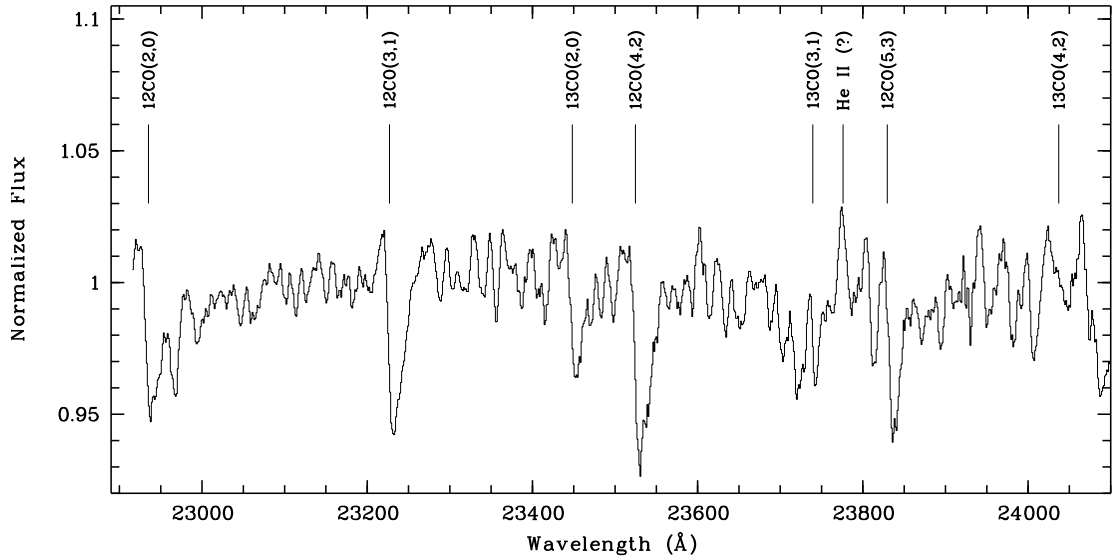,width=15.cm,angle=270,%
           bbllx=4.7cm,bblly=3.cm,bburx=17.0cm,bbury=27.cm,clip=}}\par
    \caption[irsp]{Mean K band spectrum of GRS 1915+105. It was obtained
     at the ESO VLT-Antu telescope, using
    the short wavelength (0.9--2.5 $\mu$m) arm of ISAAC,
   equipped with a 1024$\times$1024 pixel Rockwell HgCdTe array
   with an image scale of 0\farcs147/pixel.
   Using the medium resolution grating (1.2 \AA/pixel in the K band)
   yields a spectral resolution of $\sim$3000 with a 1\asec\ slit.
 Science exposures of GRS 1915+105 consisted of eight 250 sec individual 
exposures which were dithered along the slit by $\pm$10\asec.
In order to correct for atmospheric absorption, the nearby star HD 179913
(A0\,V) was observed either before or after each science exposure.
The initial data reduction steps like bias subtraction,
flatfielding and co-adding
were performed within the {\em Eclipse} package \cite{de2000}.
The extraction and wavelength calibration was done using an optimal extraction
routine within the MIDAS package.
The spectrum shown here is the sum of 5 exposures with 167 min. 
      total integration time.
CO bandheads are clearly discovered and marked by vertical lines. 
The presence of the $^{13}$CO isotope
and the equivalent width ratio of $^{12}$CO to $^{13}$CO suggests a
classification of the donor as a late-type giant.
The observed small width 
and faintness of the CO bandheads imply that the donor 
contributes only a few percent to the total K band brightness
 (see \cite{gcm01} for details) of the binary system GRS 1915+105.
         }
    \label{irsp}
\end{figure}

\begin{figure}
      \vbox{\psfig{figure=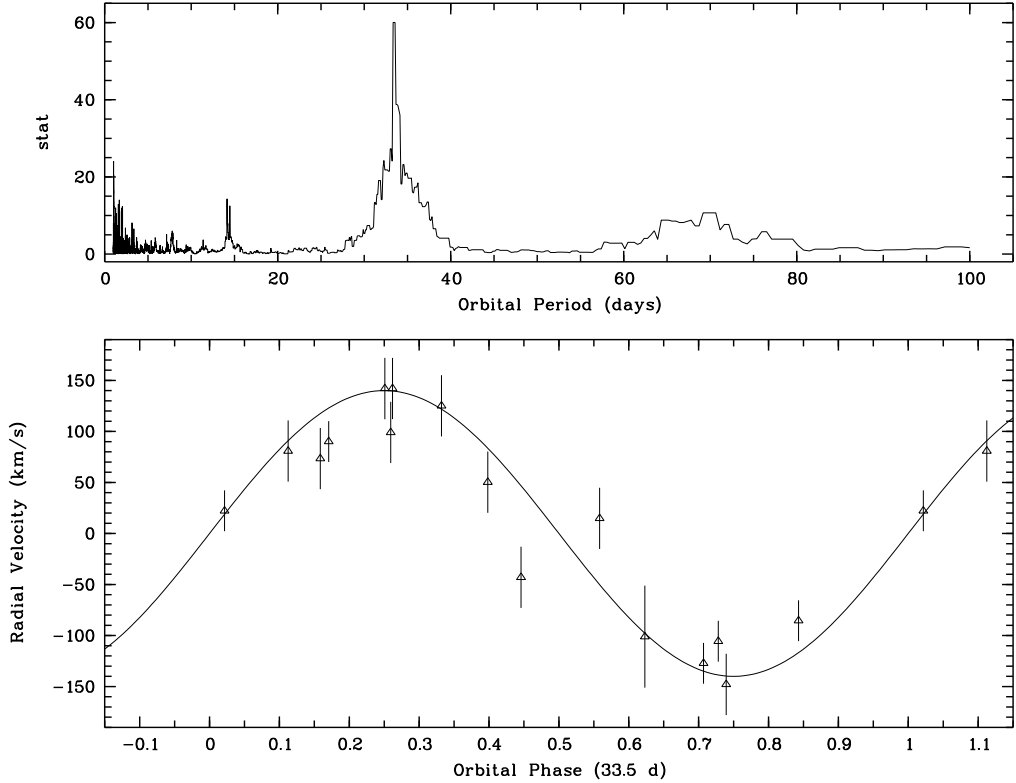,width=16.cm,angle=270}}
    \caption[rv]{Result of the period analysis of the 
         velocity variation of the four CO bandheads.
   Radial velocities were measured for the individual spectra by 
   cross-correlation of the major CO band heads, using as template a spectrum 
   of the K2\,III star HD 202135 taken with the same setting.
   {\bf Top:} Scargle periodogram after heliocentric correction of the
   individual measurements.
   {\bf Bottom:} Radial velocity curve folded over the best-fit period
   of $P_{\rm orb}$ = 33.5 days. The semi-amplitude of the velocity curve
   $K_{\rm d}$ is 140$\pm$15 km/s. Distortions of the radial velocity
   curve due to X-ray heating (e.g. \cite{psp99})
   to be unimportant because of the long orbital period.
  The systemic velocity is   $\gamma$ = --3$\pm$10 km/s which implies
  that based on the galactic rotation curve \cite{fbs89} the kinematic
  distance of GRS 1915+105 is $d = 12.1 \pm 0.8$ kpc, intermediate
  between earlier estimates \cite{fen99, mr94}.
              }
    \label{rv}
\end{figure}

\begin{figure}
    \vbox{\psfig{figure=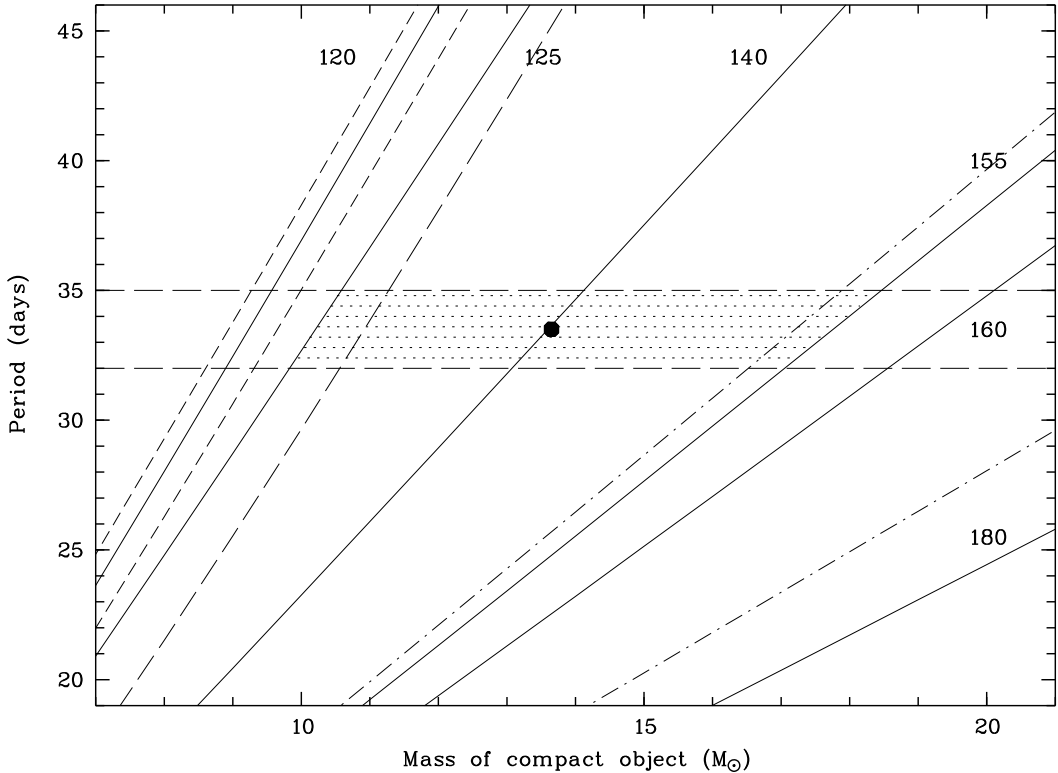,width=16.cm,angle=270}}
    \caption[mass]{Black hole mass constraints for GRS 1915+105.
 The relation of orbital period versus mass of the black hole is
plotted for various velocity amplitudes $K$ (solid lines, in km/s). We assumed
  an orbital inclination of 70\degs\ and a mass of the donor of 1.2 \msun.
The horizontal long-dashed lines indicated the boundaries of the period
uncertainty, and the radial velocity range is 125--155 km/s. Thus,
the dotted region shows the parameter space, leading to a mass of the
accreting compact object of 14$\pm$4 \msun. The implied Roche lobe size of
the donor star is 21$\pm$4 \rsun, in good agreement with the size of
a K-M giant which thus is very likely to fill its Roche lobe.
The uncertainty in the mass of the donor is shown for $K = 120$ km/s where
the slanted dashed lines correspond to 1.0, 1.4 and 2.5 \msun, respectively
(from left to right). 
While the formal uncertainty in the orbital inclination is only 2\degs,
we show the effect of relaxing the assumption of the jet being
perpendicular to the orbital plane by showing for the $K = 160$ km/s case
the corresponding curves using $i = 79$\degs\ (at which angle eclipses
would set in; left dash-dot curve) and $i = 61$\degs\ (right dash-dot curve).
Thus, when relaxing the assumptions and using the extremes, the mass
range would be 8--24 \msun.}
\label{mass}
\end{figure}

\end{document}